\documentclass[preprint,showpacs,preprintnumbers,amsmath,amssymb,aip]{revtex4-1}

\usepackage{graphicx}
\usepackage{dcolumn}
\usepackage{bm}
\usepackage{psfrag}
\usepackage[T1]{fontenc}

\begin{document}

\preprint{APS/123-QED}

\title{The time evolution equation for advective heat transport as a 
constraint for optimal bounds in Rayleigh-B\'enard convection}

\author{A. Tilgner}

\affiliation{Institute of Geophysics, University of G\"ottingen,
Friedrich-Hund-Platz 1, 37077 G\"ottingen, Germany }

\date{\today}

\begin{abstract}
Upper bounds on the heat transport and other quantities of interest in
Rayleigh-B\'enard convection are derived in previous work from constraints
resulting from the equations of time evolution for kinetic energy, the root mean
square of temperature, and the temperature averaged over horizontal planes.
Here, we investigate the effect of a new constraint derived from the time
evolution equation for the advective heat transport. This additional constraint
leads to improved bounds on the toroidal dissipation.
\end{abstract}

\pacs{47.27.te, 44.25.+f, 92.60.Fm}
\maketitle

\section{Introduction}

Thermal convection is among the best studied problems of fluid mechanics. Its
most basic idealization is Rayleigh-B\'enard convection in which a fluid fills a
plane layer infinitely extended in the horizontal directions, heated from below
and cooled from above. The applied temperature difference, or the amplitude of
the driving of the flow, is commonly expressed in terms of the Rayleigh number.
Numerical simulations and experiments can investigate convection only up to a
certain Rayleigh number. It is also possible to derive rigorous bounds on, for
example, the heat transport across the convecting layer \citep{Howard63,
Busse69,Doerin96,Plasti05,Seis15}. The bounds derived in these references are
valid at all Rayleigh numbers, but they tend to overestimate the actual heat
transport. If power laws are fitted to both the bounds and numerically computed
time averages, the fit to the bounds have a larger prefactor and usually a
larger exponent. Furthermore, the bounds do not show any Prandtl number
dependence.

These deficiencies arise because the derivation of the bounds does not exploit
the full equations of evolution, but only a few integrals deduced from them: the
energy budget, the relation between advective heat transport and dissipation,
and the temperature equation integrated over horizontal planes. Seis
\cite{Seis15} also makes use of the maximum principle for temperature. An
improvement of the bounds necessarily requires to take advantage of additional
constraints. This process is already well understood for systems of ordinary
differential equations such as the Lorenz model \cite{Tobasc18}. In the context
of convection, the implementation of further constraints is more cumbersome.
Vitanov and Busse \cite{Vitano00} split the energy budget into two equations,
one for the poloidal and one for the toroidal energy. The resulting optimization
problem preserves a dependence on the Prandtl number (and also on rotation if
one is interested in rotating convection). But there is a price to pay in this
approach. The optimization problem is not convex, and its Euler-Lagrange
equations have to be solved numerically. Because of extensive coupling between
different modes in a spectral decomposition of these equations, a numerical
solution can only be obtained at low resolution, and hence at low Rayleigh
numbers.

The approach pursued in this paper is to solve a semidefinite program (SDP).
Previously obtained bounds on Rayleigh-B\'enard convection can be reproduced by
this method \cite{Tilgne17b}. It has already been demonstrated how one can
obtain increasingly sharp bounds on solutions of systems of ordinary
differential equations by including more and more constraints
\cite{Tobasc18,Golusk17}. While the same systematic procedure is in principle possible for
the Navier-Stokes equation \cite{Cherny14}, it becomes unpractical for
Rayleigh-B\'enard convection even at modest Rayleigh numbers because all energy
unstable modes need to be retained which leads to a large SDP. Fantuzzi et al.
\cite{Fantuz18} study bounds for B\'enard-Marangoni convection and in this
context discuss in depth limitations of the SDP approach and possible future
lines of investigation. At present, the most straightforward approach remains to
formulate an optimization problem in the form of an SDP with many decoupled
linear matrix inequalities rather than a problem with a few, but large linear
matrix inequalities. 
The goal of the present paper is to derive a Prandtl number dependent 
optimization problem for Rayleigh-B\'enard
convection which is convex and whose solution does not imply a significantly
larger computational burden than optimization problems solved previously to find
bounds on various observables of interest in Rayleigh-B\'enard convection.

\section{The optimization problem}

Let us consider the problem of Rayleigh-B\'enard convection within the
Boussinesq approximation for stress free boundaries. This choice of boundary
conditions is not essential for obtaining the new constraint but it 
simplifies the calculation. A Cartesian coordinate system $(x,y,z)$ is chosen
such that a plane layer is infinitely extended in the $x-$ and $y-$ directions
and the boundaries of the layer are separated by the distance $h$ along the
$z-$direction. Gravitational acceleration $g$ is acting along the negative
$z-$direction. The layer is
filled with fluid of density $\rho$, kinematic viscosity $\nu$, thermal
diffusivity $\kappa$, and thermal expansion coefficient $\alpha$. Top and bottom
boundaries are held at the fixed temperatures $T_\mathrm{top}$ and
$T_\mathrm{top} + \Delta T$, respectively. We will consider the equations of
evolution immediately in nondimensional form, choosing for units of length,
time, and temperature deviation from $T_\mathrm{top}$ the quantities $h$,
$h^2/\kappa$ and $\Delta T$. With this choice, the equations within the
Boussinesq approximation for the fields of velocity $\bm v(\bm r,t)$,
temperature $T(\bm r,t)$ and pressure $p(\bm r,t)$ become:

\begin{eqnarray}
\frac{1}{\mathrm{Pr}} \left( \partial_t\bm v + \bm v \cdot \nabla \bm v \right)
&=& 
-\nabla p + \mathrm{Ra} \theta \bm{\hat z} + \nabla^2 \bm v
\label{eq:NS} \\
\partial_t \theta + \bm v \cdot \nabla \theta -v_z &=& \nabla^2 \theta
\label{eq:Temp} \\
\nabla \cdot \bm v &=& 0
\label{eq:conti} 
\end{eqnarray}

In these equations, $T=\theta+1-z$, so that $\theta$ represents the deviation
from the conduction profile. The Prandtl number $\mathrm{Pr}$ and the
Rayleigh number $\mathrm{Ra}$ are given by
\begin{equation}
\mathrm{Pr}=\frac{\nu}{\kappa} ~~~,~~~ 
\mathrm{Ra}=\frac{g \alpha \Delta T h^3}{\kappa \nu}
\end{equation}
and $\bm{\hat z}$ denotes the unit vector in $z-$direction.
The boundary conditions on temperature require that $\theta=0$ at $z=0$ and 1
and the stress free conditions chosen here lead to
$\partial_z v_x = \partial_z v_y = v_z = 0$ at the boundaries.

It will be helpful to reduce the number of dependent variables by introducing
poloidal and toroidal scalars $\phi$ and $\psi$ such that $\bm v = \nabla \times
\nabla \times (\phi \bm{\hat z}) + \nabla \times (\psi \bm{\hat z})$ and
$\nabla \cdot \bm v = 0$ is satisfied by construction.
The $z-$component of
the curl and the $z-$component of the curl of the curl of eq. (\ref{eq:NS})
yield the equations of evolution for $\phi$ and $\psi$,

\begin{eqnarray}
\frac{1}{\mathrm{Pr}} \left( \partial_t \nabla^2 \Delta_2 \phi
+ \bm{\hat z} \cdot \nabla \times \nabla \times \left[ (\nabla \times \bm v)
\times \bm v \right] \right)
&=&
\nabla^2 \nabla^2 \Delta_2 \phi - \mathrm{Ra} \Delta_2 \theta 
\label{eq:phi}  \\
\frac{1}{\mathrm{Pr}} \left( \partial_t  \Delta_2 \psi
- \bm{\hat z} \cdot \nabla \times \left[ (\nabla \times \bm v) \times \bm v
\right] \right)
&=&
\nabla^2 \Delta_2 \psi
\label{eq:psi} 
\end{eqnarray}

with $\Delta_2 = \partial^2_x + \partial^2_y$. For brevity, $\bm v$ is not
replaced by its expression in terms of $\phi$ and $\psi$ in these equations. The
stress free boundary conditions translate into
$\phi = \partial^2_z \phi = \partial_z \psi = 0$.

The currently known methods for obtaining upper bounds on fluid dynamic
quantities cannot exploit the equations of evolution in full detail but extract
averages from the original equations. Several types of averages will become
important: the average over
the entire volume, denoted by angular brackets without subscript,
the average over an arbitrary
plane $z=\mathrm{const.}$, denoted by $\langle ... \rangle _A$, 
and the average over time, which will be signaled by an overline.

Three useful relations between averages can now directly be obtained. The
average of eq. (\ref{eq:Temp}) over planes $z=const.$ leads to 
\begin{equation}
\partial_t \langle \theta \rangle_A =
\langle \partial_z(\theta \Delta_2 \phi)\rangle_A + \langle \partial^2_z \theta
\rangle_A
.
\label{eq:theta3}
\end{equation}
Multiplication of eq. (\ref{eq:Temp}) with
$\theta$ and a subsequent volume average yields
\begin{equation}
\partial_t \langle \frac{1}{2} \theta^2 \rangle = 
- \langle \theta \Delta_2 \phi \rangle - \langle |\nabla \theta|^2 \rangle
.
\label{eq:theta2}
\end{equation}
Finally, the dot product of $\bm v$ with eq. (\ref{eq:NS}), followed by a volume
average, leads to
\begin{equation}
\partial_t \langle \frac{1}{2} \bm v^2 \rangle =
- \mathrm{Pr} \mathrm{Ra} \langle \theta \Delta_2 \phi \rangle
- \mathrm{Pr} \langle |(\bm{\hat z}\times \nabla) \nabla^2 \phi|^2 
             + |\nabla \partial_x \psi|^2 + |\nabla \partial_y \psi|^2 \rangle
.
\label{eq:v2}
\end{equation}
These three equations form the basis of the previous work on bounds on, for
instance, the Nusselt number $\mathrm{Nu}$ given by
$\mathrm{Nu}= 1 - \langle \overline {\theta \Delta_2 \phi} \rangle
= 1+ \langle \overline{v_z \theta} \rangle$.

An additional relation will be derived below by computing 
$\partial_t \langle v_z \theta \rangle$ from the time evolution equation of the
advective heat transport $v_z \theta$. There are at least two reasons why such a
relation looks promising. Bounds have been computed for double diffusive
convection. In this problem, salinity $S$ drives convection together with
temperature, and salinity obeys the same advection-diffusion equation as
temperature except for a different diffusion constant. This problem looks
superficially identical to ordinary convection and one may think that adding the
equation for $\partial_t \langle \frac{1}{2} S^2 \rangle$ to the equations
(\ref{eq:theta3}-\ref{eq:v2}) provides us with enough information to derive
bounds on double diffusive convection. In fact, it does not. It is necessary to
include the equation for $\partial_t \langle \theta S \rangle$ to derive bounds
\cite{Balmfo06}. This demonstrates the usefulness of considering cross products
of different physical quantities and motivates us to also look at 
$\partial_t \langle v_z \theta \rangle$. Another type of flows to which bounding
methods were applied in the past are flows in periodic volumes driven by a body
force $\bm f$. \cite{Doerin02,Childr01,Rollin11,Tilgne17} In these problems, it
is essential to include the equation for 
$\partial_t \langle \bm f \cdot \bm v \rangle$ obtained by forming the dot
product of the momentum equation with $\bm f$. In Rayleigh-B\'enard convection,
the buoyancy force plays the role of the body force in the momentum equation.
From this analogy, we have another reason to look at
$\partial_t \langle v_z \theta \rangle$.

If we multiply the $z-$component of eq. (\ref{eq:NS}) by $\theta$, eq.
(\ref{eq:Temp}) by $v_z$, add the two equations and average the sum over the
volume, we obtain the time evolution equation for the advective heat transport: 
\begin{equation}
\partial_t \langle v_z \theta \rangle =
\langle v_z^2 \rangle + \mathrm{Pr} \mathrm{Ra} \langle \theta^2 \rangle
-(1+\mathrm{Pr}) \langle \nabla \theta \cdot \nabla v_z \rangle
-\langle \theta \partial_z p \rangle.
\label{eq:vz_theta}
\end{equation}
Thanks to the stress free boundary conditions and the boundary condition on
temperature, one deduces from eq. (\ref{eq:NS}) that
$\partial_z p = 0$ at the boundaries. The divergence of eq. (\ref{eq:NS}) yields
$\nabla^2 p = -\nabla \cdot [(\bm v \cdot \nabla)\bm v] + \mathrm{Pr}
\mathrm{Ra} \partial_z \theta$. It is therefore possible to split the pressure
$p$ into two terms $p_1$ and $p_2$ such that $p=p_1+p_2$ and
\begin{eqnarray}
\nabla^2 p_1 &=& -\nabla \cdot [(\bm v \cdot \nabla)\bm v]\\
\nabla^2 p_2 &=& \mathrm{Pr} \mathrm{Ra} \partial_z \theta
\end{eqnarray}
with the boundary conditions that $\partial_z p_1 = \partial_z p_2 = 0$ at
$z=0$ and 1.

It will now be shown that the expression $\langle \theta \partial_z p_2 \rangle$ is
quadratic in $\theta$ and positive. To this end, we first consider a layer with
periodic boundary conditions and finite periodicity length in the lateral
directions. At the end of the calculation, we will send the periodicity length
to infinity in order to obtain the desired result for the infinitely extended
layer. It helps to introduce two functions $f$ and $g$ defined by
$f=\partial_z p_2 /(\mathrm{Pr} \mathrm{Ra})$ (so that $\nabla^2 f =\partial_z^2
\theta$) and $\nabla^2 g=\theta$ with the boundary conditions $g=0$ at $z=0$ and
1. These definitions imply that $f$ and $g$ are periodic in $x$ and $y$ and that 
$f=\partial_z^2 g=g=0$ at $z=0$ and 1. It is possible to compute 
$\int \theta \partial_z p_2 dV$, where the integration extends over the volume
$V$ of a periodicity cell in the layer, by a sequence of integration by parts in
which the boundary integrals all vanish:
\begin{eqnarray}
\frac{1}{\mathrm{Pr} \mathrm{Ra}} \int \theta \partial_z p_2 dV & = &
\int \left( \nabla^2 g \right) f dV =
\int g \left( \nabla^2 f \right) dV =
\int g \partial_z^2 \nabla^2 g dV \nonumber \\ & = &
- \int \left( \partial_z g \right) \left( \partial_z \nabla^2 g \right) dV =
\int |\partial_z \nabla g |^2 dV.
\end{eqnarray}
We can now divide this equation by the volume $V$ and take the limit $V \rightarrow
\infty$ to conclude that
\begin{equation}
\langle \theta \partial_z p_2 \rangle = \mathrm{Pr} \mathrm{Ra}
\langle |\partial_z \nabla \Delta^{-1} \theta |^2 \rangle
\end{equation}
where $\Delta^{-1}$ denotes the inverse Laplacian for homogeneous Dirichlet
boundary conditions.

Before proceeding further, let us look at the form of the optimization problem
which yields optimal bounds from eqs. (\ref{eq:theta3}-\ref{eq:vz_theta}). There
are various ways to formulate such an optimization problem. The formulation
presented here is most closely related to the method of auxiliary functions
\cite{Cherny14,Cherny17,Tobasc18}. To facilitate the exposition of the numerical
implementation later, the formulation presented here is identical to ref.
\onlinecite{Tilgne17b}. Suppose we want to bound some objective function $Z$ which is
defined in terms of the velocity and temperature fields. We chose test functions
$\varphi_n(z)$, $n=1..N$, which depend only on $z$ and on which we project eq.
(\ref{eq:Temp}):
\begin{equation}
\partial_t \langle \varphi_n \theta \rangle = 
\langle \varphi_n \partial_z (\theta \Delta_2 \phi) \rangle
+ \langle \varphi_n \nabla^2 \theta \rangle
.
\label{eq:projection}
\end{equation}
We now construct the functional 
$F(\lambda_1 ... \lambda_N,\lambda_R,\lambda_E,\lambda_M,\theta,\phi,\psi)$ as
\begin{equation}
F(\lambda_1 ... \lambda_N,\lambda_R,\lambda_E,\lambda_M,\theta,\phi,\psi)=
\sum_{n=1}^N \lambda_n \partial_t \langle \varphi_n \theta \rangle
-\lambda_R \partial_t \langle \frac{1}{2} \theta^2 \rangle
-\frac{\lambda_E}{\mathrm{Pr}} \partial_t \langle \frac{1}{2} \bm v^2 \rangle
+\lambda_M \partial_t \langle v_z \theta \rangle 
\end{equation}
and replace all time derivatives by their expressions given in eqs.
(\ref{eq:theta2}), (\ref{eq:v2}), (\ref{eq:projection}), and
(\ref{eq:vz_theta}).
The task now is to find a set of coefficients 
$\lambda_0,\lambda_1 ... \lambda_N,\lambda_R,\lambda_E,\lambda_M$
such that the inequality
\begin{equation}
-Z + \lambda_0 + F(\lambda_1 ... \lambda_N,\lambda_R,\lambda_E,\lambda_M,\theta,\phi,\psi) \geq 0
\label{eq:p7Mitte}
\end{equation}
holds for all fields $\theta(\bm r)$, $\phi(\bm r)$ and $\psi(\bm r)$ which obey
the free slip boundary conditions and $\theta=0$ at $z=0$ and 1. If such a set
of $\lambda$'s is found, the above inequality holds in particular for the fields
taken from an actual time evolution. Taking the time average of relation (\ref{eq:p7Mitte})
thus leads to $\overline Z \leq \lambda_0 - \overline F = \lambda_0$. The time
average of $F$ is zero because it is the linear combination of time derivatives.
Replacing the time derivatives by their expressions in eqs.
(\ref{eq:theta2}), (\ref{eq:v2}), (\ref{eq:projection}), and 
(\ref{eq:vz_theta}), the best upper bound for $\overline Z$ is
given by the $\lambda_0$ which solves the optimization problem
\begin{equation}
\begin{aligned}
\text{minimize } & \lambda_0  \\
\text{subject to } & 
-Z + \lambda_0 
+\sum_{n=1}^N \lambda_n \left[ \langle \varphi_n \partial_z (\theta \Delta_2 \phi) \rangle
+ \langle \varphi_n \nabla^2 \theta \rangle \right]
+ \lambda_R \left[ \langle \theta \Delta_2 \phi \rangle + \langle |\nabla \theta|^2 \rangle \right]
\\ &
+ \lambda_E \left[ \mathrm{Ra} \langle \theta \Delta_2 \phi \rangle +
\langle |(\hat{\bm z} \times \nabla) \nabla^2 \phi|^2 \rangle + d^2 \right]
\\ &
+ \lambda_M \left[ \langle v_z^2 \rangle + \mathrm{Pr} \mathrm{Ra} \left( \langle \theta^2 \rangle - \langle |\partial_z \nabla \Delta^{-1} \theta|^2 \rangle \right)
-(1+\mathrm{Pr}) \langle \nabla \theta \cdot \nabla v_z \rangle
-\langle \theta \partial_z p_1 \rangle \right]
\geq 0 
\end{aligned}
\label{eq:opt1}
\end{equation}
with
$d^2 = \langle |\nabla \partial_x \psi|^2 + |\nabla \partial_y \psi|^2 \rangle$. 
The minimization occurs over all $\lambda$'s and the inequality needs to hold
for all $d$ and all eligible $\theta$ and $\phi$.

This is nearly in a form which leads to an SDP after a suitable discretization.
The problematic term is $\langle \theta \partial_z p_1 \rangle$ which involves a
triple product in terms of $\theta$, $\phi$ and $\psi$. However, this term is
readily reduced to a quadratic term by invoking the maximum principle which guarantees that in a
statistically stationary state, the temperature takes on values in the interval
bounded by the temperatures at the top and bottom planes of the layer, which
means that $0 \leq T = \theta+1-z \leq 1$, or equivalently,
$|T-\frac{1}{2}| \leq \frac{1}{2}$.

To make best use of the maximum principle, we want to compute 
$\langle \theta \partial_z p_1 \rangle$ in the form 
$\langle (T-\frac{1}{2}) \partial_z p_1 \rangle +
\langle (z-\frac{1}{2}) \partial_z p_1 \rangle$.
This expression can be simplified with the same artifice as before: we first
integrate over a volume $V$ with periodic boundary conditions in the horizontal
before taking the limit $V \rightarrow \infty$. The second term of the above
expression requires us to compute
\begin{equation}
\int (z-\frac{1}{2}) \partial_z p_1 dV = \int dz~ (z-\frac{1}{2}) \partial_z \int \int dx~ dy~ p_1
\label{eq:Tsdzp}
\end{equation}
in which only the horizontal average of $p_1$ appears which obeys
\begin{eqnarray}
\int \int dx~ dy~ \nabla^2 p_1 & = &
\int \int dx~ dy~ \partial_z^2 p_1 =
- \int \int dx~ dy~ \nabla \cdot [(\bm v \cdot \nabla) \bm v] 
\nonumber \\
 & = &
- \partial_z \int \int dx~ dy~ \nabla (\bm v v_z) =
- \partial_z^2 \int \int dx~ dy~ v_z^2.
\end{eqnarray}
We can thus solve
$\partial_z^2 \int \int dx~ dy~ \partial_z p_1
= - \partial_z^2 \int \int dx~ dy~ v_z^2$ 
with the boundary conditions that $\int \int dx~ dy~ \partial_z p_1 = 0$ at
$z=0$ and 1 to find
$\int \int dx~ dy~ \partial_z p_1 = -\int \int dx~ dy~ v_z^2$ since $v_z=0$ at
$z=0$ and 1. Inserting this into eq. (\ref{eq:Tsdzp}) yields
\begin{equation}
\int (z-\frac{1}{2}) \partial_z p_1 dV = - \int dz~ (z-\frac{1}{2}) \partial_z \int \int dx~ dy~ v_z^2 = \int v_z^2 dV
\label{eq:zdzp1}
\end{equation}
after another integration by parts.

The only triple product left in problem (\ref{eq:opt1}) is
$\lambda_M \langle \theta \partial_z p_1 \rangle$. This can be reduced to a
quadratic term only at the price of an inequality: 
\begin{equation}
\lambda_M \int (T-\frac{1}{2}) \partial_z p_1 dV \leq
|\lambda_M| \int |T-\frac{1}{2}| \cdot |\partial_z p_1| dV \leq
\frac{1}{2} |\lambda_M| \int |\partial_z p_1| dV
\label{eq:lambda_abs}
\end{equation}
The last expression is quadratic in $\bm v$ and could be combined with 
(\ref{eq:opt1}) to obtain a convex optimization problem which can be represented
as an SDP for numerical purposes. However, the solution of this problem would be
very expensive. The success of the previous applications of semidefinite
programming to Rayleigh-B\'enard convection \cite{Tilgne17b} relied on the fact
that if the dependence in $x$ and $y$ is represented as Fourier series, the
equations decouple in the wavenumber of the Fourier modes and only
a small number of amplitudes of Fourier modes needs to be taken into account in
a numerical computation. The term $\int |\partial_z p_1| dV$ unfortunately
destroys that decoupling.

The decoupling can be restored at the expense of a further inequality. It is
shown in the appendix that
\begin{equation}
\langle |\partial_z p_1| \rangle \leq 
\frac{1}{2} \langle \sum_{i,j} (\partial_i v_j)^2 \rangle.
\label{eq:AppA}
\end{equation}
Inserting this into eq. (\ref{eq:lambda_abs}), and (\ref{eq:lambda_abs}) together
with (\ref{eq:zdzp1}) into (\ref{eq:opt1}), leads to the following optimization
problem for the optimal bound $\lambda_0$ of the objective function
$\overline Z$, where the minimum is sought over all $\lambda$'s:
\begin{equation}
\begin{aligned}
\text{minimize } & \lambda_0  \\
\text{subject to } & 
-Z + \lambda_0 
+\sum_{n=1}^N \lambda_n \left[ \langle \varphi_n \partial_z (\theta \Delta_2 \phi) \rangle
+ \langle \varphi_n \nabla^2 \theta \rangle \right]
+ \lambda_R \left[ \langle \theta \Delta_2 \phi \rangle + \langle |\nabla \theta|^2 \rangle \right]  
\\ &
+ \lambda_E \left[ \mathrm{Ra} \langle \theta \Delta_2 \phi \rangle +
\langle |(\hat{\bm z} \times \nabla) \nabla^2 \phi|^2 \rangle + d^2 \right]  
\\ &
+ \lambda_M \left[ \mathrm{Pr} \mathrm{Ra} \left( \langle \theta^2 \rangle - \langle |\partial_z \nabla \Delta^{-1} \theta|^2 \rangle \right)
-(1+\mathrm{Pr}) \langle \nabla \theta \cdot \nabla v_z \rangle \right]
\geq \frac{1}{4} \lambda_{\mathrm{abs}}  \langle \sum_{i,j} (\partial_i v_j)^2 \rangle \\
 & 
\lambda_{\mathrm{abs}} \geq \lambda_M \geq -\lambda_{\mathrm{abs}}
\end{aligned}
\label{eq:opt2}
\end{equation}
The last line implies $|\lambda_M| \leq \lambda_{\mathrm{abs}}$. Since the
right hand side of the first inequality constraint in (\ref{eq:opt2}) is
positive, $\lambda_0$ is smallest when $\lambda_{\mathrm{abs}}$ is chosen as
small as possible, which implies that $\lambda_{\mathrm{abs}}=|\lambda_M|$ at
optimum. The velocity field is left as a variable in (\ref{eq:opt2}) so that it
is easier to trace the various terms to the preceding development, but
eventually, $v_z$ is replaced by $v_z=-(\partial_x^2+\partial_y^2)\phi$ and
$$\langle \sum_{i,j} (\partial_i v_j)^2 \rangle =
\langle |(\bm{\hat z}\times \nabla) \nabla^2 \phi|^2 \rangle + d^2$$.

Problem (\ref{eq:opt2}) contains eq. (\ref{eq:vz_theta}) as a constraint which
keeps $\mathrm{Pr}$ as a parameter in the optimization. The effect of
$\mathrm{Pr}$ is expected to be strongest for $\mathrm{Pr} \rightarrow \infty$. It
will be of interest to also study a reduced optimization problem which results
from (\ref{eq:opt2}) by introducing $\lambda_M'=\lambda_M \mathrm{Pr}$,
$\lambda_{\mathrm{abs}}' = \lambda_{\mathrm{abs}} \mathrm{Pr}$. The dissipation
is known to be bounded uniformly in $\mathrm{Pr}$, \cite{Howard63,Busse69,Doerin96} so that in the
limit $\mathrm{Pr} \rightarrow \infty$, the right hand side of the inequality
constraint disappears and we are left with a simpler problem:
\begin{equation}
\begin{aligned}
\text{minimize } & \lambda_0  \\
\text{subject to } & 
- Z + \lambda_0 
+\sum_{n=1}^N \lambda_n \left[ \langle \varphi_n \partial_z (\theta \Delta_2 \phi) \rangle
+ \langle \varphi_n \nabla^2 \theta \rangle \right]
+ \lambda_R \left[ \langle \theta \Delta_2 \phi \rangle + \langle |\nabla \theta|^2 \rangle \right]
\\ &
+ \lambda_E \left[ \mathrm{Ra} \langle \theta \Delta_2 \phi \rangle +
\langle |(\hat{\bm z} \times \nabla) \nabla^2 \phi|^2 \rangle + d^2 \right]
\\ &
+ \lambda'_M \left[ \mathrm{Ra} \left( \langle \theta^2 \rangle - \langle |\partial_z \nabla \Delta^{-1} \theta|^2 \rangle \right)
- \langle \nabla \theta \cdot \nabla v_z \rangle \right]
\geq 0
\end{aligned}
\label{eq:opt3}
\end{equation}
The virtue of this formulation is that it is independent of any sloppiness that
may have eased the derivation of eq. (\ref{eq:AppA}). The optimization
(\ref{eq:opt3}) would not be changed by a sharper inequality than
(\ref{eq:AppA}).

The numerical solution of problems (\ref{eq:opt2}) and (\ref{eq:opt3}) proceeds
in exactly the same way as in ref. \onlinecite{Tilgne17b} so that a brief summary
will suffice here. The variables $\phi$ and $\theta$ are decomposed into $N$
Chebychev polynomials $T_n$ for the $z-$direction and into plane waves in $x$
and $y$, as for example in
\begin{equation}
\theta = \sum_{n=1}^N \sum_{k_x} \sum_{k_y} {\hat \theta}_{n,k_x,k_y}
T_n(2z-1) e^{i(k_x x + k_y y)}
.
\label{eq:expansion}
\end{equation}
The test functions $\varphi_n$ are chosen as delta functions centered at the
collocation points $z_n$ defined by
\begin{eqnarray}
z_n &=& \frac{1}{2} \left[ 1+\cos \left( \pi \frac{n-1}{N-1} \right) \right]
~~~,~~~ n=1 ... N   \label{eq:colloc}\\
\varphi_n(z) &=& \delta(z-z_n)
.
\end{eqnarray}
Inserting all this into the optimization problems transforms the inequality
constraints into the condition that some symmetric matrix be positive
semidefinite. This is the standard form of an SDP. The constraints decouple in
$k^2=k_x^2+k_y^2$. Only a small number of wavenumbers actually constrain the
solution. This set of wavenumbers is determined automatically \cite{Tilgne17b}.
The matrix occurring in the SDP would be much larger if the constraints did not
decouple in $k$, hence the effort put into obtaining eq. (\ref{eq:AppA}). Some
improvements of the basic method accelerate the computation but are not
essential. These include a partial integration of eq. (\ref{eq:projection}) and
the exploitation of symmetry in $z$. \cite{Tilgne17b} The resulting SDP was
solved with the package cvxopt.

\begin{figure}
\includegraphics[width=8cm]{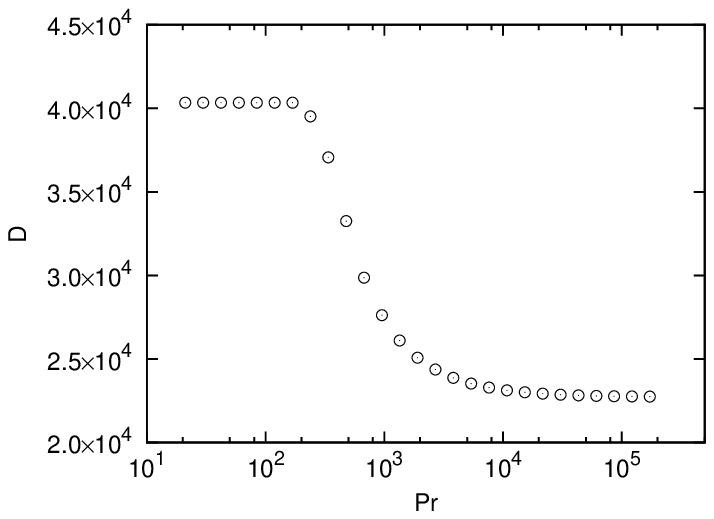}
\includegraphics[width=8cm]{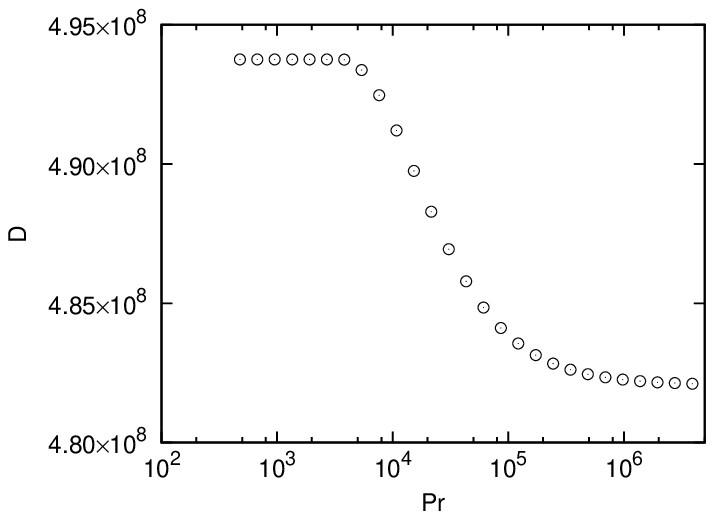}
\caption{
The bound $D$ as a function of $\mathrm{Pr}$ for $\mathrm{Ra}=1.6\times 10^4$
(left panel) and $8.192\times 10^6$ (right panel).}
\label{fig:scan}
\end{figure}

\begin{figure}
\includegraphics[width=8cm]{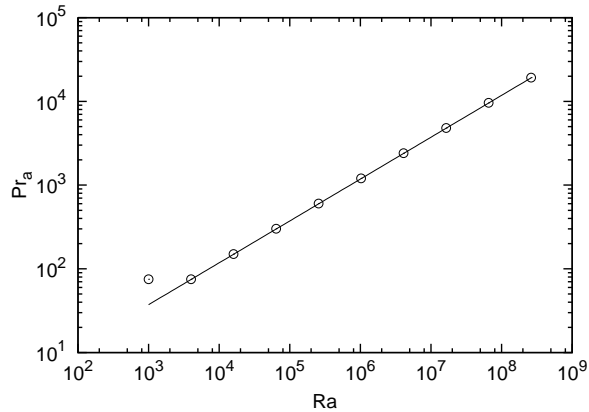}
\caption{
The Prandtl number $\mathrm{Pr}_a$ that has to be exceeded for constraint
(\ref{eq:vz_theta}) to reduce the bound $D$, plotted as function of
$\mathrm{Ra}$. The straight line is given by $1.18 \times \mathrm{Ra}^{1/2}$.}
\label{fig:Pr_active}
\end{figure}

\section{Results}

The optimization problem (\ref{eq:opt2}) distinguishes itself from previous
problems by the terms multiplied by $\lambda_M$ and $\lambda_{\mathrm{abs}}$,
which include all the terms containing $\mathrm{Pr}$. In order to test the
power of the new constraint, it seems best to choose an objective function which
varies dramatically as a function of $\mathrm{Pr}$. The dissipation of the non
poloidal components fulfills this criterion, because the flow is purely poloidal
at infinite $\mathrm{Pr}$, whereas it contains both poloidal and toroidal
components at finite $\mathrm{Pr}$. The choice $Z=d^2$ thus promises to be a
gratifying objective function.

Call $D$ the optimal $\lambda_0$ of problem (\ref{eq:opt2}) for $Z=d^2$. Fig.
\ref{fig:scan} shows $D$ as a function of $\mathrm{Pr}$ for different
$\mathrm{Ra}$. It is seen that $D$ is constant for low $\mathrm{Pr}$ until it
monotonically decreases as a function of $\mathrm{Pr}$ and asymptotes towards a
value different from zero at high $\mathrm{Pr}$. The optimization problem
(\ref{eq:opt2}) thus is not powerful enough to show that the toroidal
dissipation disappears at infinite $\mathrm{Pr}$. Let us denote with $D_0$ the
value of $D$ at small $\mathrm{Pr}$ and with $D_\infty$ the limiting value of
$D$ as $\mathrm{Pr}$ tends to infinity. $D_0$ and $D_\infty$ are functions of
$\mathrm{Pr}$.

$D_0$ is equal to the upper bound one obtains without the constraint derived
from $\partial_t \langle v_z \theta \rangle$, or with
$\lambda_M=\lambda_{\mathrm{abs}}=0$ in (\ref{eq:opt2}). This bound is the same
as the one computed in ref. \onlinecite{Tilgne17b} and obeys approximately
$D_0=0.021 \times \mathrm{Ra}^{3/2}$ at high $\mathrm{Ra}$. The new constraint
introduced in this paper becomes active and reduces the upper bound for 
$\mathrm{Pr} > \mathrm{Pr}_a$. From plots like fig. \ref{fig:scan} one finds
$\mathrm{Pr}_a$ as a function of $\mathrm{Ra}$, which is shown in fig.
\ref{fig:Pr_active}. $\mathrm{Pr}$ needs to be larger than 
$\mathrm{Pr}_a = 1.18 \times \mathrm{Ra}^{1/2}$ to obtain improved bounds.

\begin{figure}
\includegraphics[width=8cm]{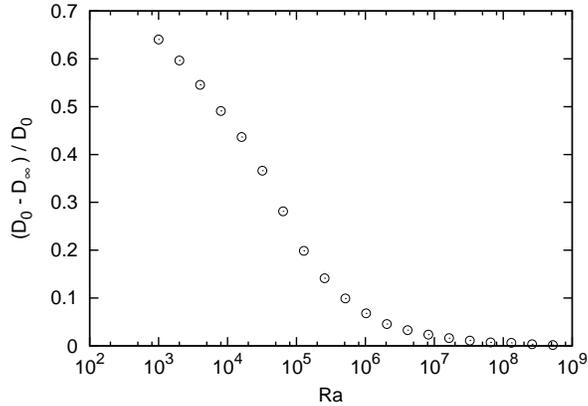}
\caption{
$(D_0-D_\infty)/D_0$ as a function of $\mathrm{Ra}$.}
\label{fig:comp_Dnp}
\end{figure}

Just as $D_0$ can be computed from a simplified optimization problem, so can
$D_\infty$ be computed from the reduced problem (\ref{eq:opt3}) which has the
advantage that it does not depend on $\mathrm{Pr}$ any more, and it is
independent of how sharp the estimate in (\ref{eq:AppA}) is. 
$D_\infty(\mathrm{Ra})$ is the lowest bound found for any $\mathrm{Pr}$ at a
given $\mathrm{Ra}$, so that $(D_0-D_\infty)/D_0$ is a measure of the maximum
fractional improvement of the bound due to the additional constraint
(\ref{eq:vz_theta}) at the given $\mathrm{Ra}$. 
This fraction is shown in fig. \ref{fig:comp_Dnp}. As can be
seen from this figure, the constraint (\ref{eq:vz_theta}) can improve the bound
by more than a factor of 2 at small $\mathrm{Ra}$, but dishearteningly, the
improvement vanishes for large $\mathrm{Ra}$. Bounding methods should be useful
at high $\mathrm{Ra}$ when ordinary time integrations become too expensive. But
it is precisely in this limit that the newly added constraint does not improve
the previously known bound.

The objective function most often considered in the context of optimum theory is
$Z = \langle v_z \theta \rangle$, which is the Nusselt number minus one. With
this choice, the fractional improvement of previous known bounds is found to be
less than $2 \times 10^{-3}$, and since this is less than the tolerances and errors
in the numerical procedure, the improvement could be exactly zero. The potential
improvement is at any rate so small that it was not considered worthwhile to
determine it accurately, or to show that it vanishes.

\section{Conclusion}

Previous work has derived bounds on the Nusselt number or other flow quantities
in convection from the time evolution equations for 
$\langle \theta \rangle _A$, $\langle \frac{1}{2} \theta^2 \rangle$, and
$\langle \frac{1}{2} \bm v^2 \rangle$. The present paper adds 
$\langle v_z \theta \rangle$ to the list. This additional constraint does not
improve the bounds on the Nusselt number by an unambiguously detectable amount.
The bounds on the toroidal dissipation on the other hand can be improved by more
than $50\%$, but there is no improvement at large $\mathrm{Ra}$. This fact is
remarkable because
it contradicts the behavior one may intuit from other results in the
optimum theory of turbulence. In order to derive bounds for flows in 3D periodic
boxes driven by a body force $\bm f$, it is necessary to take into account the
time evolution equation for $\langle \bm f \cdot \bm v \rangle$. For convection
within the Boussinesq approximation, the momentum equation contains a driving
term proportional to $\theta \bm{\hat z}$. The term analogous to 
$\langle \bm f \cdot \bm v \rangle$ of the 3D periodic box is therefore
$\langle v_z \theta \rangle$. For an immediate analogy with previous work, the
forcing should be solenoidal, so that we should first consider 
$\langle \bm v \cdot (\theta \bm{\hat z} - \nabla {\tilde p}) \rangle$, where
${\tilde p}$ is chosen such that $\nabla \cdot (\theta \bm{\hat z} - \nabla
{\tilde p}) = 0$. However, for a solenoidal velocity field with zero normal
component at the boundaries, the expression
$\langle \bm v \cdot (\theta \bm{\hat z} - \nabla {\tilde p}) \rangle$
reduces to $\langle v_z \theta \rangle$. It is therefore surprising that 
there is not a more striking improvement of bounds for convective flows if the
time evolution equation of $\langle v_z \theta \rangle$ is included in the
optimization problem. At finite $\mathrm{Pr}$, the bounding problem relies on
inequality (\ref{eq:AppA}) which can possibly be improved. In the limit of large
$\mathrm{Pr}$, however, the results are independent of this inequality and no
improvement of the results presented here can be achieved without adding yet
another constraint.

\appendix*
\section{}

The goal of this appendix is to prove eq. (\ref{eq:AppA}). To this end, we have
to compute $\frac{1}{V} \int |\partial_z p_1| dV$, where $V$ is an arbitrarily
large volume in the plane layer, and $p_1$ is given by
\begin{equation}
\nabla^2 p_1 = q
\label{eq:App1}
\end{equation}
with $\partial_z p_1=0$ at $z=0$ and 1. The variable $q$ is used as an
abbreviation for
$q = -\nabla \cdot [(\bm v \cdot \nabla)\bm v]$.
We will proceed by finding the Green's function $G(\bm r, \bm r')$ such that
$\nabla^2 G(\bm r, \bm r') = \delta(\bm r - \bm r')$
so that
$p_1(\bm r) = \int G(\bm r, \bm r') q(\bm r') d^3\bm r'$
where the integral extends over the horizontally infinitely extended layer. Once
the Green's function is known, we can estimate the desired integral from
\begin{eqnarray}
\frac{1}{V} \int |\partial_z p_1| dV & = &
\frac{1}{V} \int d^3\bm r |\int d^3 \bm r' \partial_z G(\bm r, \bm r') q(\bm r')| \leq
\frac{1}{V} \int d^3\bm r \int d^3 \bm r' |\partial_z G(\bm r, \bm r')| \cdot |q(\bm r')| \nonumber \\
 & \leq & 
\frac{1}{V} \int d^3\bm r' |q(\bm r')| \cdot 
 \max_{\bm r'} \int d^3 \bm r |\partial_z G(\bm r, \bm r')|
\end{eqnarray}
and taking the limit $V \rightarrow \infty$:

\begin{equation}
\langle |\partial_z p_1| \rangle \leq \langle |q| \rangle \cdot 
 \max_{\bm r'} \int d^3 \bm r |\partial_z G(\bm r, \bm r')|.
\label{eq:App4}
\end{equation}
Because of the translational invariance, the last factor in
eq. (\ref{eq:App4}) simplifies to
\begin{equation}
 \max_{z'} \int d^3 \bm r |\partial_z G(\bm r, \begin{pmatrix} 0 \\ 0 \\ z'
\end{pmatrix})|.
\label{eq:App2}
\end{equation}
The Green's function for eq. (\ref{eq:App1}) with Neumann boundary conditions
describes for example the potential flow out of a point source in a plane layer.
The maximization in expression (\ref{eq:App2}) asks for the distance $z'$ from
the lower boundary of the layer at which one has to place the point source so
that the absolute value of the $z-$component of the velocity integrated over the
entire layer is maximum. One intuitively expects that the maximum is reached
when the point source is located on one of the boundaries of the layer. It seems
very likely that the Green's function for exactly this problem is given
somewhere in the existing literature, but I was not able to find a suitable
reference. However, the Green's function for eq. (\ref{eq:App1}) with the
Dirichlet boundary conditions $p_1=0$ at $z=0$ and 1 describes the electrostatic
potential of a point charge between two parallel infinitely extended metallic
plates kept at zero potential, and the Green's function for this electrostatic
problem can be found in the textbook by Jackson \cite{Jackso99}. Adapting the
result in this textbook from Dirichtlet to Neumann boundary conditions leads
to
\begin{equation}
G(\bm r, \begin{pmatrix} 0 \\ 0 \\ z' \end{pmatrix}) =
\frac{1}{2\pi} \ln s -\frac{1}{\pi} \sum_{n=1}^\infty K_0(n\pi s) \cos(n\pi z) \cos(n\pi z') + C
\label{eq:App3}
\end{equation}
where the position $\bm r$ is given in cylindrical coordinates $(s,\varphi,z)$
and $C$ is a constant left unspecified by the Neumann boundary conditions and
which is irrelevant because we are only interested in $\partial_z G$. $K_0$ is a
modified Bessel function of the second kind in the usual notation.

\begin{figure}
\includegraphics[width=8cm]{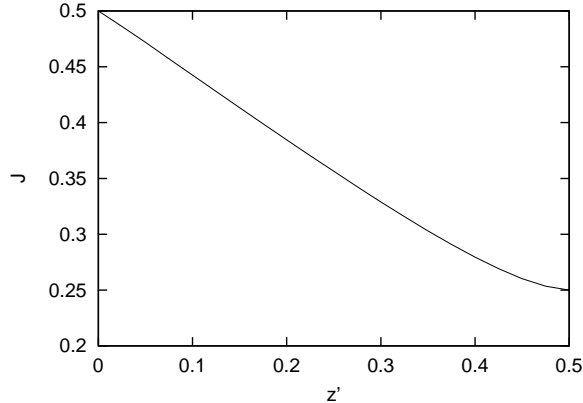}
\caption{The integral $J=\int d^3 \bm r |\partial_z G(\bm r, \bm r')|$  as a
function of $z'$, the $z-$coordinate of $\bm r'$.}
\label{fig:potential_max}
\end{figure}

\begin{figure}
\includegraphics[width=10cm]{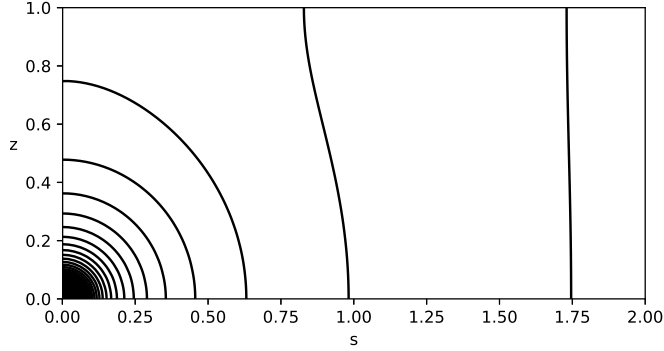}
\caption{Contour plot of $g(\bm r)$ in a cross section of the layer.}
\label{fig:Potential_plot}
\end{figure}

We will abstain from formal proofs for two properties which we will deduce from
numerical evaluation of eq. (\ref{eq:App3}) 
and which match the intuition about $G$ one may have from
its interpretation as a potential flow out of a point source. The first
observation concerns the $z'$ at which the maximum in (\ref{eq:App2}) is
realized. Fig. \ref{fig:potential_max} shows $\int d^3 \bm r |\partial_z G|$ as
a function of $z'$ computed numerically from the first 40 terms of eq.
(\ref{eq:App3}). As expected, this integral is maximal for $z'=0$ (or $z'=1$ by
symmetry). The value of the maximum is numerically close to $1/2$.

This second point can be made more precise with the help of a weaker
observation. Let us introduce for brevity $g(\bm r)$ as the Green's function
corresponding to a point source located at the origin,
$g(\bm r) = G(\bm r, \bm 0)$. A contour plot of $g(\bm r)$ in cylindrical
coordinates is shown in fig. \ref{fig:Potential_plot}. The second property to
extract from numerical calculation and which is again expected from the
interpretation of $g(\bm r)$ as potential flow, is that $\partial_z g$ has the
same sign everywhere. This second property is important because it implies that
we can avail ourselves of the absolute values in expression (\ref{eq:App2}). Using the formulas
$\int_0^\infty s K_0(s) ds = 1$
and
$\sum_{n=1}^\infty \frac{1}{n} \sin (n\alpha) = \frac{\pi-\alpha}{2}$,
one finds
\begin{eqnarray}
\max_{z'} \int d^3 \bm r |\partial_z G(\bm r, \begin{pmatrix} 0 \\ 0 \\ z' \end{pmatrix})| & = &
\int d^3 \bm r |\partial_z g(\bm r)| =
\int d^3 \bm r \partial_z g(\bm r) 
\nonumber \\ & = &
\int_0^1 dz \sum_{n=1}^\infty \sin(n \pi z) n \int_0^\infty ds 2 \pi s K_0(n \pi s) =
\frac{1}{2}
\end{eqnarray}
so that eq. (\ref{eq:App4}) leads to
\begin{equation}
\langle |\partial_z p_1| \rangle \leq
\frac{1}{2} \langle |\nabla \cdot [(\bm v \cdot \nabla)\bm v]| \rangle.
\end{equation}
For solenoidal fields $\bm v$, this can be brought into a form which is more
pleasant for the SDP:
\begin{eqnarray}
\frac{1}{V} \int |\partial_z p_1| dV & \leq &
\frac{1}{2} \frac{1}{V} \int |\sum_{i,j} (\partial_i v_j) (\partial_j v_i)| dV \leq
\frac{1}{2} \frac{1}{V} \int \sum_{i,j} |\partial_i v_j| |\partial_j v_i| dV 
\nonumber \\
 & \leq &
\frac{1}{2} \frac{1}{V} \sum_{i,j} \sqrt{\int(\partial_i v_j)^2 dV} \sqrt{\int(\partial_j v_i)^2 dV} 
\nonumber \\
 & \leq &
\frac{1}{2} \frac{1}{V} \sum_{i,j} \frac{1}{2} \left( \int(\partial_i v_j)^2 dV + \int(\partial_j v_i)^2 dV \right)
\end{eqnarray}
which in the limit $V \rightarrow \infty$ leads to eq. (\ref{eq:AppA}).


\begin{thebibliography}{18}%
\makeatletter
\providecommand \@ifxundefined [1]{%
 \@ifx{#1\undefined}
}%
\providecommand \@ifnum [1]{%
 \ifnum #1\expandafter \@firstoftwo
 \else \expandafter \@secondoftwo
 \fi
}%
\providecommand \@ifx [1]{%
 \ifx #1\expandafter \@firstoftwo
 \else \expandafter \@secondoftwo
 \fi
}%
\providecommand \natexlab [1]{#1}%
\providecommand \enquote  [1]{``#1''}%
\providecommand \bibnamefont  [1]{#1}%
\providecommand \bibfnamefont [1]{#1}%
\providecommand \citenamefont [1]{#1}%
\providecommand \href@noop [0]{\@secondoftwo}%
\providecommand \href [0]{\begingroup \@sanitize@url \@href}%
\providecommand \@href[1]{\@@startlink{#1}\@@href}%
\providecommand \@@href[1]{\endgroup#1\@@endlink}%
\providecommand \@sanitize@url [0]{\catcode `\\12\catcode `\$12\catcode
  `\&12\catcode `\#12\catcode `\^12\catcode `\_12\catcode `\%12\relax}%
\providecommand \@@startlink[1]{}%
\providecommand \@@endlink[0]{}%
\providecommand \url  [0]{\begingroup\@sanitize@url \@url }%
\providecommand \@url [1]{\endgroup\@href {#1}{\urlprefix }}%
\providecommand \urlprefix  [0]{URL }%
\providecommand \Eprint [0]{\href }%
\providecommand \doibase [0]{http://dx.doi.org/}%
\providecommand \selectlanguage [0]{\@gobble}%
\providecommand \bibinfo  [0]{\@secondoftwo}%
\providecommand \bibfield  [0]{\@secondoftwo}%
\providecommand \translation [1]{[#1]}%
\providecommand \BibitemOpen [0]{}%
\providecommand \bibitemStop [0]{}%
\providecommand \bibitemNoStop [0]{.\EOS\space}%
\providecommand \EOS [0]{\spacefactor3000\relax}%
\providecommand \BibitemShut  [1]{\csname bibitem#1\endcsname}%
\let\auto@bib@innerbib\@empty
\bibitem [{\citenamefont {Howard}(1963)}]{Howard63}%
  \BibitemOpen
  \bibfield  {author} {\bibinfo {author} {\bibfnamefont {L.}~\bibnamefont
  {Howard}},\ }\bibfield  {title} {\enquote {\bibinfo {title} {Heat transport
  by turbulent convection},}\ }\href@noop {} {\bibfield  {journal} {\bibinfo
  {journal} {J. Fluid Mech.}\ }\textbf {\bibinfo {volume} {17}},\ \bibinfo
  {pages} {405--432} (\bibinfo {year} {1963})}\BibitemShut {NoStop}%
\bibitem [{\citenamefont {Busse}(1969)}]{Busse69}%
  \BibitemOpen
  \bibfield  {author} {\bibinfo {author} {\bibfnamefont {F.}~\bibnamefont
  {Busse}},\ }\bibfield  {title} {\enquote {\bibinfo {title} {{On Howard's
  upper bound for heat transport by turbulent convection}},}\ }\href@noop {}
  {\bibfield  {journal} {\bibinfo  {journal} {J. Fluid Mech.}\ }\textbf
  {\bibinfo {volume} {37}},\ \bibinfo {pages} {457--477} (\bibinfo {year}
  {1969})}\BibitemShut {NoStop}%
\bibitem [{\citenamefont {Doering}\ and\ \citenamefont
  {Constantin}(1996)}]{Doerin96}%
  \BibitemOpen
  \bibfield  {author} {\bibinfo {author} {\bibfnamefont {C.}~\bibnamefont
  {Doering}}\ and\ \bibinfo {author} {\bibfnamefont {P.}~\bibnamefont
  {Constantin}},\ }\bibfield  {title} {\enquote {\bibinfo {title} {{Variational
  bounds on energy dissipation in incompressible flows:III. Convection}},}\
  }\href@noop {} {\bibfield  {journal} {\bibinfo  {journal} {Phys. Rev. E}\
  }\textbf {\bibinfo {volume} {53}},\ \bibinfo {pages} {5957--5981} (\bibinfo
  {year} {1996})}\BibitemShut {NoStop}%
\bibitem [{\citenamefont {Plasting}\ and\ \citenamefont
  {Ierley}(2005)}]{Plasti05}%
  \BibitemOpen
  \bibfield  {author} {\bibinfo {author} {\bibfnamefont {C.}~\bibnamefont
  {Plasting}}\ and\ \bibinfo {author} {\bibfnamefont {G.}~\bibnamefont
  {Ierley}},\ }\bibfield  {title} {\enquote {\bibinfo {title}
  {{Infinite-Prandtl-number convection. Part 1. Conservative bounds}},}\
  }\href@noop {} {\bibfield  {journal} {\bibinfo  {journal} {J. Fluid Mech.}\
  }\textbf {\bibinfo {volume} {542}},\ \bibinfo {pages} {343--363} (\bibinfo
  {year} {2005})}\BibitemShut {NoStop}%
\bibitem [{\citenamefont {Seis}(2015)}]{Seis15}%
  \BibitemOpen
  \bibfield  {author} {\bibinfo {author} {\bibfnamefont {C.}~\bibnamefont
  {Seis}},\ }\bibfield  {title} {\enquote {\bibinfo {title} {Scaling bounds on
  dissipation in turbulent flows},}\ }\href@noop {} {\bibfield  {journal}
  {\bibinfo  {journal} {J. Fluid Mech.}\ }\textbf {\bibinfo {volume} {777}},\
  \bibinfo {pages} {591--603} (\bibinfo {year} {2015})}\BibitemShut {NoStop}%
\bibitem [{\citenamefont {Tobasco}, \citenamefont {Goluskin},\ and\
  \citenamefont {Doering}(2018)}]{Tobasc18}%
  \BibitemOpen
  \bibfield  {author} {\bibinfo {author} {\bibfnamefont {I.}~\bibnamefont
  {Tobasco}}, \bibinfo {author} {\bibfnamefont {D.}~\bibnamefont {Goluskin}}, \
  and\ \bibinfo {author} {\bibfnamefont {C.}~\bibnamefont {Doering}},\
  }\bibfield  {title} {\enquote {\bibinfo {title} {Optimal bounds and extremal
  trajectories for time averages in dynamical systems},}\ }\href@noop {}
  {\bibfield  {journal} {\bibinfo  {journal} {Phys. Lett. A}\ }\textbf
  {\bibinfo {volume} {382}},\ \bibinfo {pages} {382 -- 386} (\bibinfo {year}
  {2018})}\BibitemShut {NoStop}%
\bibitem [{\citenamefont {Vitanov}\ and\ \citenamefont
  {Busse}(2000)}]{Vitano00}%
  \BibitemOpen
  \bibfield  {author} {\bibinfo {author} {\bibfnamefont {N.}~\bibnamefont
  {Vitanov}}\ and\ \bibinfo {author} {\bibfnamefont {F.}~\bibnamefont
  {Busse}},\ }\bibfield  {title} {\enquote {\bibinfo {title} {{Bounds on the
  convective heat transport in a rotating layer}},}\ }\href@noop {} {\bibfield
  {journal} {\bibinfo  {journal} {Phys. Rev. E}\ }\textbf {\bibinfo {volume}
  {63}},\ \bibinfo {pages} {016303} (\bibinfo {year} {2000})}\BibitemShut
  {NoStop}%
\bibitem [{\citenamefont {Tilgner}(2017{\natexlab{a}})}]{Tilgne17b}%
  \BibitemOpen
  \bibfield  {author} {\bibinfo {author} {\bibfnamefont {A.}~\bibnamefont
  {Tilgner}},\ }\bibfield  {title} {\enquote {\bibinfo {title} {Bounds on
  poloidal kinetic energy in plane layer convection},}\ }\href@noop {}
  {\bibfield  {journal} {\bibinfo  {journal} {Phys. Rev. Fluids}\ }\textbf
  {\bibinfo {volume} {2}},\ \bibinfo {pages} {123502} (\bibinfo {year}
  {2017}{\natexlab{a}})}\BibitemShut {NoStop}%
\bibitem [{\citenamefont {Goluskin}(2018)}]{Golusk17}%
  \BibitemOpen
  \bibfield  {author} {\bibinfo {author} {\bibfnamefont {D.}~\bibnamefont
  {Goluskin}},\ }\bibfield  {title} {\enquote {\bibinfo {title} {{Bounding
  averages rigorously using semidefinite programming: mean moments of the
  Lorenz system}},}\ }\href@noop {} {\bibfield  {journal} {\bibinfo  {journal}
  {Journal of Nonlinear Science}\ }\textbf {\bibinfo {volume} {28}},\ \bibinfo
  {pages} {621--651} (\bibinfo {year} {2018})}\BibitemShut {NoStop}%
\bibitem [{\citenamefont {Chernyshenko}\ \emph {et~al.}(2014)\citenamefont
  {Chernyshenko}, \citenamefont {Goulart}, \citenamefont {Huang},\ and\
  \citenamefont {Papachristodoulou}}]{Cherny14}%
  \BibitemOpen
  \bibfield  {author} {\bibinfo {author} {\bibfnamefont {S.}~\bibnamefont
  {Chernyshenko}}, \bibinfo {author} {\bibfnamefont {P.}~\bibnamefont
  {Goulart}}, \bibinfo {author} {\bibfnamefont {D.}~\bibnamefont {Huang}}, \
  and\ \bibinfo {author} {\bibnamefont {Papachristodoulou}},\ }\bibfield
  {title} {\enquote {\bibinfo {title} {{Polynomial sum of squares in fluid
  dynamics: a review with a look ahead}},}\ }\href@noop {} {\bibfield
  {journal} {\bibinfo  {journal} {Phil. Trans. Roy. Soc. A}\ }\textbf {\bibinfo
  {volume} {372}},\ \bibinfo {pages} {20130350} (\bibinfo {year}
  {2014})}\BibitemShut {NoStop}%
\bibitem [{\citenamefont {Fantuzzi}, \citenamefont {Pershin},\ and\
  \citenamefont {Wynn}(2018)}]{Fantuz18}%
  \BibitemOpen
  \bibfield  {author} {\bibinfo {author} {\bibfnamefont {G.}~\bibnamefont
  {Fantuzzi}}, \bibinfo {author} {\bibfnamefont {A.}~\bibnamefont {Pershin}}, \
  and\ \bibinfo {author} {\bibfnamefont {A.}~\bibnamefont {Wynn}},\ }\bibfield
  {title} {\enquote {\bibinfo {title} {{Bounds on heat transfer for
  B\'enard-Marangoni convection at infinite Prandtl number}},}\ }\href@noop {}
  {\bibfield  {journal} {\bibinfo  {journal} {J. Fluid Mech.}\ }\textbf
  {\bibinfo {volume} {837}},\ \bibinfo {pages} {562--596} (\bibinfo {year}
  {2018})}\BibitemShut {NoStop}%
\bibitem [{\citenamefont {Balmforth}\ \emph {et~al.}(2006)\citenamefont
  {Balmforth}, \citenamefont {Ghadge}, \citenamefont {Kettapun},\ and\
  \citenamefont {Mandre}}]{Balmfo06}%
  \BibitemOpen
  \bibfield  {author} {\bibinfo {author} {\bibfnamefont {N.}~\bibnamefont
  {Balmforth}}, \bibinfo {author} {\bibfnamefont {S.}~\bibnamefont {Ghadge}},
  \bibinfo {author} {\bibfnamefont {A.}~\bibnamefont {Kettapun}}, \ and\
  \bibinfo {author} {\bibfnamefont {S.}~\bibnamefont {Mandre}},\ }\bibfield
  {title} {\enquote {\bibinfo {title} {Bounds on double-diffusive
  convection},}\ }\href@noop {} {\bibfield  {journal} {\bibinfo  {journal} {J.
  Fluid Mech.}\ }\textbf {\bibinfo {volume} {569}},\ \bibinfo {pages} {29--50}
  (\bibinfo {year} {2006})}\BibitemShut {NoStop}%
\bibitem [{\citenamefont {Doering}\ and\ \citenamefont
  {Foias}(2002)}]{Doerin02}%
  \BibitemOpen
  \bibfield  {author} {\bibinfo {author} {\bibfnamefont {C.}~\bibnamefont
  {Doering}}\ and\ \bibinfo {author} {\bibfnamefont {C.}~\bibnamefont
  {Foias}},\ }\bibfield  {title} {\enquote {\bibinfo {title} {Energy
  dissipation in body-forced turbulence},}\ }\href@noop {} {\bibfield
  {journal} {\bibinfo  {journal} {J. Fluid Mech.}\ }\textbf {\bibinfo {volume}
  {467}},\ \bibinfo {pages} {289--306} (\bibinfo {year} {2002})}\BibitemShut
  {NoStop}%
\bibitem [{\citenamefont {Childress}, \citenamefont {Kerswell},\ and\
  \citenamefont {Gilbert}(2001)}]{Childr01}%
  \BibitemOpen
  \bibfield  {author} {\bibinfo {author} {\bibfnamefont {S.}~\bibnamefont
  {Childress}}, \bibinfo {author} {\bibfnamefont {R.}~\bibnamefont {Kerswell}},
  \ and\ \bibinfo {author} {\bibfnamefont {A.}~\bibnamefont {Gilbert}},\
  }\bibfield  {title} {\enquote {\bibinfo {title} {{Bounds on dissipation for
  Navier-Stokes flow with Kolmogorov forcing}},}\ }\href@noop {} {\bibfield
  {journal} {\bibinfo  {journal} {Physica D}\ }\textbf {\bibinfo {volume}
  {158}},\ \bibinfo {pages} {105--128} (\bibinfo {year} {2001})}\BibitemShut
  {NoStop}%
\bibitem [{\citenamefont {Rollin}, \citenamefont {Dubief},\ and\ \citenamefont
  {Doering}(2011)}]{Rollin11}%
  \BibitemOpen
  \bibfield  {author} {\bibinfo {author} {\bibfnamefont {B.}~\bibnamefont
  {Rollin}}, \bibinfo {author} {\bibfnamefont {Y.}~\bibnamefont {Dubief}}, \
  and\ \bibinfo {author} {\bibfnamefont {C.}~\bibnamefont {Doering}},\
  }\bibfield  {title} {\enquote {\bibinfo {title} {{Variations on Kolmogorov
  flow: turbulent energy dissipation and mean flow profiles}},}\ }\href@noop {}
  {\bibfield  {journal} {\bibinfo  {journal} {J. Fluid Mech.}\ }\textbf
  {\bibinfo {volume} {670}},\ \bibinfo {pages} {204--213} (\bibinfo {year}
  {2011})}\BibitemShut {NoStop}%
\bibitem [{\citenamefont {Tilgner}(2017{\natexlab{b}})}]{Tilgne17}%
  \BibitemOpen
  \bibfield  {author} {\bibinfo {author} {\bibfnamefont {A.}~\bibnamefont
  {Tilgner}},\ }\bibfield  {title} {\enquote {\bibinfo {title} {{Scaling laws
  and bounds for the turbulent G.O. Roberts dynamo}},}\ }\href@noop {}
  {\bibfield  {journal} {\bibinfo  {journal} {Phys. Rev. Fluids}\ }\textbf
  {\bibinfo {volume} {2}},\ \bibinfo {pages} {024606} (\bibinfo {year}
  {2017}{\natexlab{b}})}\BibitemShut {NoStop}%
\bibitem [{\citenamefont {Chernyshenko}(2017)}]{Cherny17}%
  \BibitemOpen
  \bibfield  {author} {\bibinfo {author} {\bibfnamefont {S.}~\bibnamefont
  {Chernyshenko}},\ }\bibfield  {title} {\enquote {\bibinfo {title}
  {{Relationship between the methods of bounding time averages}},}\ }\href@noop
  {} {\bibfield  {journal} {\bibinfo  {journal} {arXiv:1704.02475}\ } (\bibinfo
  {year} {2017})}\BibitemShut {NoStop}%
\bibitem [{\citenamefont {Jackson}(1999)}]{Jackso99}%
  \BibitemOpen
  \bibfield  {author} {\bibinfo {author} {\bibfnamefont {J.}~\bibnamefont
  {Jackson}},\ }\href@noop {} {\emph {\bibinfo {title} {{Classical
  Electrodynamics, 3rd edition}}}}\ (\bibinfo  {publisher} {Wiley},\ \bibinfo
  {address} {New York},\ \bibinfo {year} {1999})\BibitemShut {NoStop}%
\end{thebibliography}

%

\end{document}